# Experiment-driven Characterization of Full-Duplex Wireless Systems

Melissa Duarte, Chris Dick and Ashutosh Sabharwal

**Abstract**

We present an experiment-based characterization of passive suppression and active self-interference cancellation mechanisms in full-duplex wireless communication systems. In particular, we consider passive suppression due to antenna separation at the same node, and active cancellation in analog and/or digital domain. First, we show that the average amount of cancellation increases for active cancellation techniques as the received self-interference power increases. Our characterization of the average cancellation as a function of the self-interference power allows us to show that for a constant signal-to-interference ratio at the receiver antenna (before any active cancellation is applied), the rate of a full-duplex link increases as the self-interference power increases. Second, we show that applying digital cancellation after analog cancellation can sometimes increase the self-interference, and thus digital cancellation is more effective when applied selectively based on measured suppression values. Third, we complete our study of the impact of self-interference cancellation mechanisms by characterizing the probability distribution of the self-interference channel before and after cancellation.

## I. Introduction

Current deployed wireless communication systems employ either a time-division or frequency-division approach to bidirectional communication. This requires dividing the temporal and/or spectral resources into orthogonal resources and results in half-duplex wireless communication systems. The key deterrent in implementing a full-duplex wireless communication system, which

This work was partially supported by NSF Grants CNS-0551692, CNS-0923479 and CNS-1012921. The first author was also supported by a Xilinx Fellowship and a Roberto Rocca Fellowship.

M. Duarte and A. Sabharwal are with the Department of Electrical and Computer Engineering, Rice University, Houston, TX, 77005 USA, e-mail: {mduarte, ashu}@rice.edu.

C. Dick is with Xilinx Inc., San Jose, CA, 95124 USA, e-mail: chris.dick@xilinx.com.

consists on same band simultaneous bidirectional communication, is the large self-interference from a node's own transmissions in comparison to the signal of interest from the distant transmitting antenna. The large self-interference spans most of the dynamic range of the analog-to-digital converter in the receiving chain, which in turn dramatically increases the quantization noise for the signal of interest. However, recent experimental results for indoor scenarios have shown that it is possible to implement self-interference cancellation mechanisms that can sufficiently attenuate the self-interference such that the resulting full-duplex wireless system can achieve higher rates than a half-duplex system [1, 2]. Hence, recent results have demonstrated that full-duplex systems are a feasible option for future indoor wireless communications.

In this paper, we perform a data-driven analysis of the full-duplex architecture proposed in [2]. Based on extensive experimental data, our main contribution consists in characterizing the impact of different self-interference cancellation mechanisms on the performance of full-duplex wireless communication systems. We consider three methods to reduce self-interference, which can be classified as either *passive* or *active*. The passive suppression is simply attenuation caused by path-loss made possible by *antenna separation* between the transmitting antenna and the receiving antenna on the same node. To reduce the dynamic range of the self-interference, we use the active *analog cancellation* proposed in [2] where an additional RF chain is used to cancel the self-interference at the receiving antenna, before the analog-to-digital converter. In addition, we also study active *digital cancellation*, where the self-interference is removed in baseband after analog to digital conversion. Our characterization of total and individual cancellation mechanisms, based on extensive experimentations, shows that a total average cancellation of 74 dB can be achieved.

Our results show that the average amount of self-interference cancelled by active cancellation increases as the power of the self-interference signal increases. This result is intuitive because the canceller relies on estimating the self-interference channel, and a higher self-interference power implies lower channel estimation error and hence better cancellation. Our characterization of active cancellation as a function of the self-interference power allows us to show that for a constant signal to interference ratio (SIR) at the receiver antenna (this is the SIR before active cancellation), the rate of a full-duplex link increases as the self-interference power increases. This appears counter-intuitive at first but follows from the fact that the average amount of self-interference cancellation increases with increasing self-interference power. Related work on implementation of self-interference cancellation mechanisms [1–5] has reported measured values of the amount of cancellation that can be achieved but a characterization of the effect of increasing self-interference power on the amount of active cancellation and rate has not been reported before. Authors in [5] have argued that the rate performance of full-duplex is independent of the transmitter power, this conclusion is valid for the particular model and implementation in [5], however we demonstrate that this doesn't generalize to all full-duplex implementations. Specifically, we demonstrate an implementation where the full-duplex rate increases as the transmit power increases.

Related work [1–4] has shown that digital cancellation, while by itself insufficient to deal with self-interference, can increase the total amount of cancellation when applied after analog cancellation. However, intuitively it is clear that in an ideal scenario where analog cancellation could achieve infinite dB of cancellation then having digital cancellation after analog cancellation would be unnecessary. This leads to the natural question regarding when is digital cancellation useful. We present results that show that the self-interference suppression achieved by digital cancellation when applied after analog cancellation decreases as the self-interference suppression achieved by analog cancellation increases. Further, our results show that when analog cancellation achieves large suppression then applying digital cancellation after analog cancellation can increase the noise in the system. No previous work on full-duplex system implementation has analyzed the performance of digital cancellation as a function of the performance of analog cancellation. Our results show that digital cancellation is an excellent "safety net," i.e, for the cases when analog cancellation delivers poor suppression, digital cancellation is most useful when applied selectively frame-by-frame based on measured suppression performance.

We complete our study of the full-duplex architecture in [2] by characterizing the distribution of the self-interference channel before and after active cancellation. Before applying active cancellation the self-interference channel is the channel between two antennas that are close to each other, consequently there is a strong Line-Of-Sight (LOS) component and the magnitude of the self-interference channel can be modeled as a Ricean distribution with large $K$-factor. After applying active cancellation the strong LOS component is reduced, hence, the magnitude of the self-interference channel can be modeled as a Ricean distribution with smaller $K$-factor [6]. We present a characterization of the $K$-factor values before and after active cancellation, such characterization has not been reported before for any full-duplex architecture.

Full-duplex wireless communications have been considered in the context of two-way (bidirectional) communication [1–6] and in the context of full-duplex relays [7–9]. Our analysis of the amount of self-interference cancellation achieved by the full-duplex architecture in [2] is applicable to full-duplex nodes either in a relay or a two-way communication scenario. Our analysis of the rate performance of full-duplex will focus only on a two-way system.

The rest of the paper is organized as follows. In Section II-A we derive equations that model the full-duplex architecture we have implemented and serve as a theoretical framework that we will use for explanation of our results. A description of the experiment setup is presented in Section III. In Section IV we present a characterization of the amount self-interference cancellation achieved and a characterization of the $K$-factor of the self-interference channel before and after active cancellation. In Section V we present an analysis of the achievable rate performance of full-duplex two-way systems. Conclusions are presented in Section VI.

## II. Channel Model for Full-duplex

Fig. 1 shows the block diagram for the narrowband full-duplex node architecture we consider. Part of the signal processing is implemented using the following major blocks: upsampling and pulse shaping (UPS), matched filter and downsampling (MFD), digital to analog converters (DACs), analog to digital converters (ADCs), and transmitter and receiver radios. The transmit radios (Tx Radio) upconvert the input signal from baseband (BB) to radio frequency (RF) and the receive radio (Rx Radio) downconverts from RF to BB. In this section we first present self-interference channel models with different stages of passive and active suppression. At the end of the section we present the channel model for a two-way full-duplex system.

### A. Channel Model With Antenna Separation (No Active Cancellation)

In Fig. 1, signal $x_i[n,f]$ denotes the $n$-th symbol transmitted from Node $i$ during frame $f$. A frame consists of $N_{\text{sym}}$ consecutive transmitted symbols. We define $x_i[n,f] = \sqrt{E_{\text{S},i}} s_i[n,f]$ where $s_i[n,f]$ is the transmitted constellation symbol normalized to unit energy and $E_{\text{S},i}$ denotes the average symbol energy. Consequently $E[|s_i[n,f]|^2] = 1$ and $E[|x_i[n,f]|^2] = E_{\text{S},i}$. We use $h_{\text{I},i}[f]$ to denote the wireless self-interference channel at Node $i$ and we model this channel as a random variable that remains constant during the transmission of a frame $f$ and changes from one frame to the next. We define $\Omega_{\text{I},i} = E[|h_{\text{I},i}[f]|^2]$.

Antenna separation is the simplest passive self-interference suppression mechanism and the amount of cancellation achieved by antenna separation depends on the propagation loss of the signal traveling through wireless channel $h_{\mathrm{I},i}[f]$. At Node $i$ the received self-interference signal after antenna separation is equal to $h_{\mathrm{I},i}[f]x_{\mathrm{RF},i}(t)$.

### *B. With Analog Cancellation*

The analog self-interference cancellation we consider, depicted in Fig. 1, consists in adding a cancelling signal to the received signal in the analog domain. The hardware components used for the implementation of analog cancellation consist of one DAC, one transmitter radio, one RF attenuator, and one RF adder. The output of the RF attenuator is connected to the RF adder via a wire. At Node $i$ the canceling signal input to the RF adder is equal to $h_{\mathrm{Z},i}[f]c_{\mathrm{RF},i}(t) = -h_{\mathrm{Z},i}[f]\widehat{\kappa}_{\mathrm{AC},i}[f]x_{\mathrm{RF},i}(t)$, where $h_{\mathrm{Z},i}[f]$ denotes the magnitude and phase change that affect signal $\widehat{\kappa}_{\mathrm{AC},i}[f]x_{\mathrm{RF},i}(t)$ when passing through an attenuator and a wire to the RF adder. Analog self-interference cancellation uses $\widehat{\kappa}_{\mathrm{AC},i}[f]$ such that $h_{\mathrm{I},i}[f]x_{\mathrm{RF},i}(t) - h_{\mathrm{Z},i}[f]\widehat{\kappa}_{\mathrm{AC},i}[f]x_{\mathrm{RF},i}(t) = 0$. One can easily see that if $\widehat{\kappa}_{\mathrm{AC},i}[f] = h_{\mathrm{I},i}[f]/h_{\mathrm{Z},i}[f]$ then the self-interference at the input of Node $i$'s receiver radio will be completely cancelled. However, the estimate of channels $h_{\mathrm{I},i}[f]$ and $h_{\mathrm{Z},i}[f]$ is not perfect due to additive noise and other distortions in the system. We define the noiseless cancellation coefficient used for analog self-interference cancellation during frame $f$ at Node $i$ as $\kappa_{\mathrm{AC},i}[f] = h_{\mathrm{I},i}[f]/h_{\mathrm{Z},i}[f]$ and we use $\widehat{\kappa}_{\mathrm{AC},i}[f]$ to denote the noisy estimate of $\kappa_{\mathrm{AC},i}[f]$. The self-interference at Node $i$ after applying analog cancellation is equal to

$$y_{\mathrm{I},i}^{\mathrm{AC}}[n,f] = \left(h_{\mathrm{I},i}[f] - h_{\mathrm{Z},i}[f]\widehat{\kappa}_{\mathrm{AC},}[f]\right) x_i[n,f]. \tag{1}$$

If digital cancellation (described below in Section II-C) is not being used then $\widehat{\kappa}_{\mathrm{DC},i}[f] = 0$ and the received self-interference $y_{\mathrm{I},i}[n,f]$ is equal to $y_{\mathrm{I},i}^{\mathrm{AC}}[n,f]$.

We note that any gains applied by the radios are included in the energy term $E_{\mathrm{S},i}$ and the cancellation coefficient $\widehat{\kappa}_{\mathrm{AC},i}[f]$. This reduces the amount of notation required to present the model. We also note that the additional transmitter radio used for analog cancellation does not require a power amplifier since the signal used for analog cancellation is being transmitted over a wire. For our specific implementation, the radio used for analog cancellation had a power amplifier by default so we used an RF attenuator connected in series, as shown in Fig. 1, in order to cancel the effect of the power amplifier. The attenuator used was a passive device with a fixed 35 dB of attenuation. The RF adder used [10] was also a passive device.

### C. With Analog and Digital Cancellation

Digital cancellation at Node $i$ is simply adding $-\widehat{\kappa}_{\mathrm{DC},i}[f]x_i[n,f]$ in baseband to the received signal. Since digital cancellation without analog cancellation does not yield an interesting system [1, 2] (else full-duplex would not be a challenge), we will not consider the case of digital cancellation without analog cancellation. From (1) we observe that perfect digital cancellation of the self-interference signal at Node $i$ can be achieved by setting $\widehat{\kappa}_{\mathrm{DC},i}[f] = h_{\mathrm{I},i}[f] - h_{\mathrm{Z},i}[f]\widehat{\kappa}_{\mathrm{AC},}[f]$, however, Node $i$ will not have a perfect estimate of $h_{\mathrm{I},i}[f] - h_{\mathrm{Z},i}[f]\widehat{\kappa}_{\mathrm{AC},}[f]$ due to noise and other distortions present in the system. We define the noiseless cancellation coefficient used for digital self-interference cancellation at Node $i$ as $\kappa_{\mathrm{DC},i}[f] = h_{\mathrm{I},i}[f] - h_{\mathrm{Z},i}[f]\widehat{\kappa}_{\mathrm{AC},i}[f]$ and we use $\widehat{\kappa}_{\mathrm{DC},i}[f]$ to denote the noisy estimate of $\kappa_{\mathrm{DC},i}[f]$. The self-interference after analog and digital cancellation is equal to

$$y_{\mathrm{I},i}^{\mathrm{ACDC}}[n,f] = (h_{\mathrm{I},i}[f] - h_{\mathrm{Z},i}[f]\widehat{\kappa}_{\mathrm{AC},i}[f] - \widehat{\kappa}_{\mathrm{DC},i}[f])x_i[n,f]. \tag{2}$$

Hence, $y_{\mathrm{I},i}[n,f] = y_{\mathrm{I},i}^{\mathrm{ACDC}}[n,f]$ when both analog and digital cancellation are applied.

### D. Notation Simplification and Summary of Self-Interference Model Parameters

We now rewrite (1) and (2) in terms of the average amount of cancellation achieved by different self-interference cancellation mechanisms and in terms of the normalized self-interference channel after cancellation. This will allow us to write the equations for the received self-interference as function of a few parameters and will reduce the notation. We define the average amount of cancellation achieved by a self-interference cancellation mechanism as the ratio of the self-interference energy before cancellation to the self-interference energy after cancellation. Hence, the average amount of cancellation achieved by analog cancellation at Node $i$, $\alpha_{\mathrm{AC},i}$, is given by

$$\alpha_{\mathrm{AC},i} = \frac{E\left[|h_{\mathrm{I},i}[f]x_i[n,f]|^2\right]}{E\left[|(h_{\mathrm{I},i}[f] - h_{\mathrm{Z},i}[f]\widehat{\kappa}_{\mathrm{AC},i}[f])\,x_i[n,f]|^2\right]} = \frac{\Omega_{\mathrm{I},i}}{E\left[|h_{\mathrm{I},i}[f] - h_{\mathrm{Z},i}[f]\widehat{\kappa}_{\mathrm{AC},i}[f]|^2\right]} \tag{3}$$

and the average amount of cancellation achieved by combined analog and digital cancellation at Node $i$, $\alpha_{\mathrm{ACDC},i}$ is given by

$$\begin{aligned} \alpha_{\mathrm{ACDC}_i} &= \left(E\left[|h_{\mathrm{I},i}[f]x_i[n,f]|^2\right]\right)\Big/\left(E\left[|(h_{\mathrm{I},i}[f] - h_{\mathrm{Z},i}[f]\widehat{\kappa}_{\mathrm{AC},i}[f] - \widehat{\kappa}_{\mathrm{DC},i}[f])\,x_i[n,f]|^2\right]\right) \\ &= \Omega_{\mathrm{I},i}\Big/\left(E\left[|h_{\mathrm{I},i}[f] - h_{\mathrm{Z},i}[f]\widehat{\kappa}_{\mathrm{AC},i}[f] - \widehat{\kappa}_{\mathrm{DC},i}[f]|^2\right]\right) \end{aligned} \tag{4}$$

Further, we define the self-interference channel after cancellation as the coefficient that multiplies the self-interference signal after applying self-interference cancellation. From (1) we have that the self-interference channel after analog cancellation at Node $i$ is equal to $h_{\mathrm{I},i}[f] - h_{\mathrm{Z},1}[f]\widehat{\kappa}_{\mathrm{AC},1}[f]$. From (2) we have that the self-interference channel after analog and digital cancellation at Node $i$ is equal to $h_{\mathrm{I},i}[f] - h_{\mathrm{Z},i}[f]\widehat{\kappa}_{\mathrm{AC},i}[f] - \widehat{\kappa}_{\mathrm{DC},i}[f]$. We define the normalized self-interference channel after analog cancellation at Node $i$ as

$$h_{\mathrm{AC},i}[f] = \left(h_{\mathrm{I},i}[f] - h_{\mathrm{Z},i}[f]\widehat{\kappa}_{\mathrm{AC},i}[f]\right) \Big/ \left(\sqrt{E\left[|h_{\mathrm{I},i}[f] - h_{\mathrm{Z},i}[f]\widehat{\kappa}_{\mathrm{AC},i}[f]|^2\right]}\right) \tag{5}$$

and the normalized self-interference channel after analog and digital cancellation at Node $i$ as

$$\begin{aligned} h_{\mathrm{ACDC},i}[f] &= \left(h_{\mathrm{I},i}[f] - h_{\mathrm{Z},i}[f]\widehat{\kappa}_{\mathrm{AC},i}[f] - \widehat{\kappa}_{\mathrm{DC},i}[f]\right) \Big/ \\ &\quad \left(\sqrt{E\left[|h_{\mathrm{I},i}[f] - h_{\mathrm{Z},i}[f]\widehat{\kappa}_{\mathrm{AC},i}[f] - \widehat{\kappa}_{\mathrm{DC},i}[f]|^2\right]}\right). \end{aligned} \tag{6}$$

We rewrite (1) and (2) in terms of $\alpha_{\mathrm{AC},i}$, $\alpha_{\mathrm{ACDC},i}$, $h_{\mathrm{AC},i}[f]$, and $h_{\mathrm{ACDC},i}[f]$ as follows. For a full-duplex system with active self-interference cancellation mechanism $\Phi$ the self-interference at Node $i$ after active cancellation is given by

$$y^{\Phi}_{\mathrm{I},i}[n,f] = h_{\Phi,i}[f]\sqrt{\Omega_{\mathrm{I},i}/\alpha_{\Phi,i}}\,x_i[n,f], \tag{7}$$

where $\Phi \in \{\mathrm{AC}, \mathrm{ACDC}\}$; recall we use AC to denote Analog Cancellation and ACDC to denote Analog and Digital Cancellation. From (7) we observe that the self interference at Node $i$ depends on $h_{\Phi,i}[f]$, $\Omega_{\mathrm{I},i}$, $\alpha_{\Phi,i}$, and $x_i[n,f]$. An experiment-based characterization of $h_{\Phi,i}[f]$, $\Omega_{\mathrm{I},i}$, and $\alpha_{\Phi,i}$ for our full-duplex implementation will be presented in Section IV.

### *E. Channel Model for Two-way Full-duplex*

Our rate analysis will focus only on full-duplex two-way communication where two nodes, Node 1 and Node 2 both implemented using the architecture in Fig. 1, have data for each other and transmit and receive simultaneously in the same frequency band. In a two-way full-duplex system with active self-interference cancellation $\Phi$ the received signal at Node 1 after active cancellation is given by $y_{\Phi,1}[n,f] = h_{\mathrm{S},2}x_2[n,f] + y^{\Phi}_{\mathrm{I},1}[n,f] + w_1[n,f]$ and the received signal at Node 2 after active cancellation is given by $y_{\Phi,2}[n,f] = h_{\mathrm{S},1}x_1[n,f] + y^{\Phi}_{\mathrm{I},2}[n,f] + w_2[n,f]$. We use $h_{\mathrm{S},1}$ to denote the wireless channel between the single transmitter antenna at Node 1 and the single receiver antenna at Node 2, $h_{\mathrm{S},2}$ to denote the wireless channel between the single transmitter antenna at Node 2 and the single receiver antenna at Node 1, and $w_i[n,f]$ to denote the white Gaussian noise added at Node $i$'s receiver. We define $\Omega_{\mathrm{S},i} = E[|h_{\mathrm{S},i}[f]|^2]$ and $N_0 = E[|w_1[n,f]|^2]$. The average received SINR per symbol at Node 1 is given by

$$\mathrm{SINR}_{\Phi,1} = \frac{E\left[|h_{\mathrm{S},2}[f]x_2[n,f]|^2\right]}{E\left[\left|h_{\Phi,1}[f]\sqrt{\Omega_{\mathrm{I},1}/\alpha_{\Phi,1}}x_1[n,f]+w_1[n,f]\right|^2\right]} = \frac{\Omega_{\mathrm{S},2}E_{\mathrm{S},2}}{\frac{\Omega_{\mathrm{I},1}E_{\mathrm{S},1}}{\alpha_{\Phi,1}}+N_0} = \frac{1}{\frac{1}{\alpha_{\Phi,1}\mathrm{SIR}_{\mathrm{AS},1}}+\frac{1}{\mathrm{SNR}_1}} \quad (8)$$

where $\mathrm{SNR}_1 = \Omega_{\mathrm{S},2}E_{\mathrm{S},2}/N_0$ is the average received signal to noise ratio (SNR) per symbol at Node 1 and $\mathrm{SIR}_{\mathrm{AS},1} = (\Omega_{\mathrm{S},2}E_{\mathrm{S},2})/(\Omega_{\mathrm{I},1}E_{\mathrm{S},1})$ is the average received signal to interference ratio (SIR) per symbol at Node 1 with only antenna separation and before applying any active cancellation. From (8) we observe that using active self-interference cancellation mechanism $\Phi$ improves the SIR by a factor of $\alpha_{\Phi,1}$ compared to the SIR after antenna separation and before applying active cancellation. The average received SINR per symbol at Node 2 can be obtained by swapping subindices 1 and 2 in (8).

We note that the domain in the expectations $E[\cdot]$ in equations (3)-(6) and (8) is all the set of random variables in the argument. The random variables in our equations are $h_{\mathrm{I},i}$, $h_{\mathrm{Z},i}$, $h_{\mathrm{S},i}$, $x_i$, $\widehat{\kappa}_{\mathrm{AC},i}$, $\widehat{\kappa}_{\mathrm{DC},i}$, and $w_i$.

## III. Experiment Setup and Scenarios Considered

We used the WARPLab framework [11] to implement two full-duplex nodes, Node 1 and Node 2, both with the architecture shown in Fig. 1. Two WARP nodes corresponding to Nodes 1 and 2 were connected via an Ethernet switch to a host PC running MATLAB. The digital baseband waveforms (samples) input to the DACs were constructed in MATLAB and downloaded from the MATLAB workspace to transmit buffers in the FPGA of the WARP nodes. Transmission and reception of over-the-air signals was done in real-time using the WARP hardware. The samples at the output of the ADCs were stored in receiver buffers and loaded to the MATLAB workspace on the host PC and processing of these samples was done in MATLAB.

We now explain in more detail the way in which $\widehat{\kappa}_{\mathrm{AC},i}[f]$ was computed in our full-duplex implementation. We observed from experiment data that channels $h_{\mathrm{I},i}[f]$ and $h_{\mathrm{Z},i}[f]$ varied very slowly from one frame to the next. This slow variation was expected since channel $h_{\mathrm{Z},i}[f]$ is a wire channel and channel $h_{\mathrm{I},i}[f]$ is the wireless channel between two antennas that are fixed and at a small distance $d$ from each other. Consequently, the computation of $\widehat{\kappa}_{\mathrm{AC},i}[f]$ in our full-duplex implementation took advantage of the slow variation of $h_{\mathrm{I},i}[f]$ and $h_{\mathrm{Z},i}[f]$ by averaging their ten most recent estimates. Specifically, in our experiments $\widehat{\kappa}_{\mathrm{AC},i}[f]$ was computed as $\widehat{\kappa}_{\mathrm{AC},i}[f] = \frac{\sum_{l=0}^{L-1} \frac{1}{L}\widehat{h}_{\mathrm{I},i}[f-l]}{\sum_{l=0}^{L-1} \frac{1}{L}\widehat{h}_{\mathrm{Z},i}[f-l]}$ where $L = 10$, $\widehat{h}_{\mathrm{I},i}[f]$ denotes the noisy estimate of $h_{\mathrm{I},i}[f]$, and $\widehat{h}_{\mathrm{Z},i}[f]$ denotes the noisy estimate of $h_{\mathrm{Z},i}[f]$. Estimates $\widehat{h}_{\mathrm{I},i}[f]$ and $\widehat{h}_{\mathrm{Z},i}[f]$ were computed using training sent in every frame. We did experiments where we observed that computation of $\widehat{\kappa}_{\mathrm{AC},i}[f]$ using $L = 10$ resulted in larger average amount of cancellation compared to using only $L = 1$. We also performed experiments where we observed that increasing $L$ from 10 to 15 did not result in a measurable improvement of the average amount of cancellation achieved by analog cancellation. Hence, we decided to set $L = 10$. Characterizing the coherence time of the self-interference channel $h_{\mathrm{I},i}[f]$ and the effect of different channel estimators on the amount of cancellation achieved by analog cancellation is important for the design of optimal training for computation of the cancellation coefficient. This is left as part of future work and for this paper the computation of $\widehat{\kappa}_{\mathrm{AC},i}[f]$ is based on averaging $\widehat{h}_{\mathrm{I},i}[f]$ and $\widehat{h}_{\mathrm{Z},i}[f]$ over ten frames as explained above. For the implementation of digital cancellation our computation of $\widehat{\kappa}_{\mathrm{DC},i}[f]$ uses only the most recent estimate of $h_{\mathrm{I},i}[f] - h_{\mathrm{Z},i}[f]\widehat{\kappa}_{\mathrm{AC},}[f]$ because the value of $h_{\mathrm{I},i}[f] - h_{\mathrm{Z},i}[f]\widehat{\kappa}_{\mathrm{AC},}[f]$ depends on errors in the estimate of $h_{\mathrm{I},i}[f]$ and $h_{\mathrm{Z},i}[f]$ and these estimation errors vary randomly from one frame to the next. The time diagram of a frame for our two-way full-duplex implementation is shown in Fig. 2. The time between two consecutive frames is equal to 25 s, this duration is mainly due to the latency of the non real time processing and the reading and writing of samples between the MATLAB workspace and the FPGA buffers.

In our experiments both nodes were located at a height of 2 meters above the floor with a Line-Of-Sight (LOS) between all antennas. The distance between nodes was fixed to $D = 8.5$ m. For the separation between same-node antennas we used values of $d \in \{10\text{ cm}, 20\text{ cm}, 40\text{ cm}\}$. The antennas used were 2.4 GHz 7 dBi Desktop Omni [12]. We used transmission power values of $\mathrm{P_T} \in \{0\text{ dBm}, 5\text{ dBm}, 10\text{ dBm}, 15\text{ dBm}\}$. We ran the experiments in the laboratory of the Center for Multimedia Communication at Rice University, the laboratory is located in the second floor of Duncan Hall building. We ran the experiments during school holiday recess, hence, there were few people walking in the lab and our experiment setup corresponds to a low mobility scenario. The carrier frequency was centered at 2.4 GHz. The sampling frequency of ADCs and DACs was equal to 40 MHz, the ADCs had 14 bits of resolution, the DACs had 16 bits of resolution, and we implemented a single subcarrier narrowband system with bandwidth of 0.625 MHz (128 samples per symbol). Our self-interference cancellation can be extended to wideband (e.g. 802.11g) by applying the proposed cancellation schemes per subcarrier, as shown in [13]. Transmitted QPSK symbols were shaped using a squared root raised cosine pulse shaping filter with roll-off factor equal to one. For our experiments, the number of symbols per frame was equal to 100 ($N_{\text{sym}} = 100$) and the number of frames transmitted in one experiment was equal to 800 ($N_{\text{frames}} = 800$). The value of $N_{\text{sym}}$ was limited to 100 due to the constraint on the total number of samples that can be stored per WARPLab buffer ($2^{14}$). Setting $N_{\text{frames}} = 800$ allowed us to approximate that during an experiment the channels conditions remained approximately constant and variations were only due to small scale variations.

## IV. Measurement-based Characterization of Channel Parameters

### A. Average Cancellation by Active Cancellations

In our experiments, we used measurements of signal power before and after active cancellation in order to estimate $\alpha_{\text{AC},i}$ and $\alpha_{\text{ACDC},i}$ for each Node $i = 1, 2$ as follows. For a fixed inter-antenna distance $d$ at a node, and transmit power $\text{P}_{\text{T}}$, we measured the power of the received self-interference signal at Node $i$ during frame $f$, which we label as $\text{P}_{\text{RI},i}[f]$. This measured power is the power of the self-interference simply due to path loss from antenna separation and before applying active cancellation. We also measured the power of the self-interference signal after analog cancellation at Node $i$ during frame $f$, which we label as $\text{P}_{\text{AC},i}[f]$, and the power of the self-interference signal after combined analog and digital cancellation at Node $i$ during frame $f$, which we label as $\text{P}_{\text{ACDC},i}[f]$. Per frame measurements of signal power were used to obtain estimates of average power by averaging over all frames. Specifically, the average power of the received self-interference signal was computed as $\text{P}_{\text{RI},i} = \sum_{f=1}^{N_{\text{frames}}} \text{P}_{\text{RI},i}[f]$, the average power of the self-interference signal after analog cancellation was computed as $\text{P}_{\text{AC},i} = \sum_{f=1}^{N_{\text{frames}}} \text{P}_{\text{AC},i}[f]$, and the average power of the self-interference signal after analog and digital cancellation was computed as $\text{P}_{\text{ACDC},i} = \sum_{f=1}^{N_{\text{frames}}} \text{P}_{\text{ACDC},i}[f]$. Using these average powers we computed the value of $\alpha_{\Phi}$, for $\Phi \in \{\text{AC, ACDC}\}$, at Node $i$ as $\alpha_{\Phi,i}$ (dB) $= \text{P}_{\text{RI},i}$ (dBm) $-$ $\text{P}_{\Phi,i}$ (dBm). The values of $\alpha_{\text{AC},i}$ and $\alpha_{\text{ACDC},i}$ from experimental data are shown in Fig. 3(a) and (b), respectively, as a function of $\text{P}_{\text{RI},i}$. Each marker in Fig. 3 corresponds to an 800 frame experiment for a fixed $d$, $\text{P}_{\text{T}}$, and $i$. Since we considered three values for $d$, four values for $\text{P}_{\text{T}}$, and two values for $i$, a total of 24 different experiment scenarios were considered. For each scenario we ran two 800 frame experiments resulting in 48 data points from experiment results which are all shown in Fig. 3. We do not specify the values of $d$ and $i$ in Fig. 3 in order to avoid cluttering the figures. Varying either $d$, $\mathrm{P_T}$ or $i$ will result in a different value of $\mathrm{P}_{\mathrm{RI},i}$ and we analyze the dependence of $\alpha_{\mathrm{AC},i}$ and $\alpha_{\mathrm{ACDC},i}$ on $\mathrm{P}_{\mathrm{RI},i}$.

To understand the dominant trends of the experiment data in Fig. 3 we compute the *constant fit*, $\alpha^{\mathrm{con}}_{\Phi,i}$, and *linear fit*, $\alpha^{\mathrm{lin}}_{\Phi,i}$, to the data. The constant and linear fits, both computed using a least squares fit, are shown in Fig. 3. The key observation is that the linear fit captures the behavior of $\alpha$ as a function received interference power better than the constant fit. For example, for values of $\mathrm{P}_{\mathrm{RI},i}$ lower than -40 dBm the constant fit $\alpha^{\mathrm{con}}_{\Phi,i}$ tends to overestimate the values of $\alpha_{\Phi,i}$ and for values of $\mathrm{P}_{\mathrm{RI},i}$ larger than -25 dBm the constant fit $\alpha^{\mathrm{con}}_{\Phi,i}$ tends to underestimate the value of $\alpha_{\Phi,i}$. The better fit between the experiment data and the linear fit leads to the following first result.

*Result 1:* As the average power of the received self-interference, $\mathrm{P}_{\mathrm{RI},i}$, increases, the average amount of interference suppression achieved by active cancellation (both AC and ACDC) also increases.

*Reasons for Result 1*: In order to implement the active cancellation mechanisms, we first need to estimate the wireless self-interference channel. The average power of the signal used to estimate the wireless self-interference channel is $\mathrm{P}_{\mathrm{RI},i}$. As $\mathrm{P}_{\mathrm{RI},i}$ increases, the error in the estimation of the wireless self-interference channel decreases, and thus, the cancellation process is more exact leading to larger suppression of self-interference. ■

Result 1 captures the total average performance of self-interference cancellation for both $\Phi = \{\mathrm{AC, ACDC}\}$. We dig deeper into the relative contributions of analog and digital cancellation in ACDC and discover the following result.

*Result 2:* As the average performance of analog cancellation gets better, the average effectiveness of digital cancellation after analog cancellation reduces.

In Fig. 4, we plot the average amount of cancellation achieved by digital cancellation after analog cancellation, computed as $\alpha_{\mathrm{DC},i}$ (dB) $= \alpha_{\mathrm{ACDC},i}$ (dB) $- \alpha_{\mathrm{AC},i}$ (dB), as a function of the average amount of cancellation achieved by analog cancellation. Results in Fig. 4 show that, in agreement with Result 2, as $\alpha_{\mathrm{AC},i}$ increases $\alpha_{\mathrm{DC},i}$ decreases. In Fig. 4 we also show the value of $\alpha_{\mathrm{DC},i}$ computed based on the constant fit and the linear fit. Results in Fig. 4 show that the dominant behavior is again better captured by the linear fit.

Dual to Result 2 is the following result.

*Result 3:* The smaller the amount of suppression achieved by analog cancellation during a frame, the larger the probability that applying digital cancellation after analog cancellation will result in an increase of the total suppression during that frame.

In contrast to the average system performance in Results 1 and 2, Result 3 relates to frame-by-frame performance. In our experiments, the suppression achieved by analog cancellation during frame $f$ was computed as $\alpha_{\mathrm{AC},i}[f]$ (dB) $=$ $\mathrm{P}_{\mathrm{RI},i}[f]$ (dBm) $-$ $\mathrm{P}_{\mathrm{AC},i}[f]$ (dBm) and the suppression achieved by analog and digital cancellation during frame $f$ was computed as $\alpha_{\mathrm{ACDC},i}[f]$ (dB) $=$ $\mathrm{P}_{\mathrm{RI},i}[f]$ (dBm) $-$ $\mathrm{P}_{\mathrm{ACDC},i}[f]$ (dBm). The suppression achieved by digital cancellation after analog cancellation during frame $f$ was computed as $\alpha_{\mathrm{DC},i}[f] = \alpha_{\mathrm{ACDC},i}[f] - \alpha_{\mathrm{AC},i}[f]$. Digital cancellation resulted in an increase in total suppression during frame $f$ if $\alpha_{\mathrm{DC},i}[f]$ (dB) $> 0$.

Result 3 is verified by experiment results in Fig. 5 which show the probability that digital cancellation results in an increase in total suppression during a frame as a function of the suppression achieved by analog cancellation during a frame. For example, for values $\alpha_{\mathrm{AC},i}[f]$ between 24 dB and 25 dB the probability of having $\alpha_{\mathrm{DC},i}[f]$ (dB) $> 0$ is equal to 95 %. We observe from Fig. 5 that digital cancellation after analog cancellation becomes more effective as $\alpha_{\mathrm{AC},i}[f]$ decreases. Fig. 5 shows that digital cancellation is effective for the frames where analog cancellation achieves less than 32 dB of suppression, since in these frames the probability that digital cancellation increases the total suppression is greater than 50 %. However, for frames where analog cancellation achieves more than 32 dB of cancellation, applying digital cancellation after analog cancellation is not effective since the probability that digital cancellation results in an increase of the total amount of cancellation is less than 50 % hence it is most likely that digital cancellation will increase the self-interference. Based on Result 3 we propose a design rule for full duplex systems which will be presented in Section V-C.

*Reasons for Result 2 and Result 3*: Intuitively it is clear that if analog cancellation can achieve perfect cancellation (infinite dB of cancellation) then digital cancellation is unnecessary. In fact, if analog cancellation can achieve perfect cancellation then applying digital cancellation can result in an increase in the self-interference. This can be observed from (2). Notice that in case of perfect analog cancellation we have that $h_{\mathrm{I},i}[f] - h_{\mathrm{Z},i}[f]\widehat{\kappa}_{\mathrm{AC},i}[f] = 0$ but due to noise in the system the value of $\widehat{\kappa}_{\mathrm{DC},i}[f]$ will not be equal to zero and will correspond to a measurement of noise hence adding $-\widehat{\kappa}_{\mathrm{DC},i}[f]x_i[n,f]$ to the signal after analog cancellation will result in an increase in the self-interference. We observe that as the performance of analog cancellation improves, the noise in the estimation of $h_{\mathrm{I},i}[f] - h_{\mathrm{Z},i}[f]\widehat{\kappa}_{\mathrm{AC},i}[f]$ increases since $h_{\mathrm{I},i}[f] - h_{\mathrm{Z},i}[f]\widehat{\kappa}_{\mathrm{AC},i}[f]$ becomes a smaller quantity and this reduces the effectiveness of applying digital cancellation after analog cancellation. Although our implementation of analog cancellation does not achieve perfect cancellation we do observe from experiment results in Fig. 5 that for values of $\alpha_{\mathrm{AC},i}[f]$ larger than 32 dB it is most likely that applying digital cancellation will increase the self-interference. This is consistent with average performance results in Fig. 4 which show that when the average cancellation achieved by analog cancellation is larger than 30 dB then digital cancellation after analog cancellation can result in an increase in the average self-interference which results in negative values of $\alpha_{\mathrm{DC},i}$. ■

### B. Passive Suppression and Total Cancellation

The amount of passive suppression at Node $i$ was computed as $\Omega_{\mathrm{I},i}$ (dB) $=$ $\mathrm{P_T}$ (dBm) $-$ $\mathrm{P}_{\mathrm{RI},i}$ (dBm). Averaging all the measurements of $\Omega_{\mathrm{I},i}$ for a fixed $d$ we obtain that the average passive suppression for our implementation is equal to 34 dB for $d = 10$ cm, 41 dB for $d = 20$ cm, and 44 dB for $d = 40$ cm. As expected, larger $d$ results in larger passive suppression.

The total cancellation (cancellation achieved by combining passive suppression, active analog cancellation, and active digital cancellation) at Node $i$ was computed as $\alpha_{\mathrm{TOT},i}$ (dB) $=$ $\mathrm{P_T}$ (dBm) $-$ $\mathrm{P}_{\mathrm{ACDC},i}$ (dBm). Averaging all the measurements of $\alpha_{\mathrm{TOT},i}$ for a fixed $d$ we obtain that the average total cancellation for our implementation is equal to 66 dB for $d = 10$ cm, 73 dB for $d = 20$ cm, and 74 dB for $d = 40$ cm. The total cancellation is observed to increase with increasing $d$. For the transmit powers we have used in our experiments (between 0 dBm and 15 dBm), a total cancellation of 74 dB will not bring the self-interference down to the noise floor, which is approximately -90 dBm. However, as we will show in Section V-B, an average cancellation of 74 dB can result in full-duplex rates larger than half-duplex rates.

### C. Characterizing the $K$-factor

We assume that the self-interference channel has a Ricean distribution and we model $|h_{\mathrm{I},i}[f]|$ and $|h_{\Phi,i}[f]|$ as Ricean with $K$-factor $K_{\mathrm{I},i}$ and $K_{\Phi,i}$ respectively. Estimates of $K_{\mathrm{I},i}$ and $K_{\Phi,i}$ were computed based on experiment data and the moment based estimator presented in equation (3) of [14]. Each value of $K_{\mathrm{I},i}$, $K_{\Phi,i}$ that we computed was based on 800 consecutive measurements of $|h_{\mathrm{I},i}[f]|$ and $|h_{\Phi,i}[f]|$ respectively made at a fixed $d$, and $\mathrm{P_T}$. We note that the variations in

each set of 800 consecutive measurements corresponded to small scale variations hence each set of 800 measurements can be used to obtain one estimate of the $K$-factor.

Our assumption that the self-interference channel is Ricean distributed makes sense intuitively because before applying active cancellation the self-interference channel is the channel between two antennas that are close to each other, hence, there is a strong LOS component and the effect of active cancellation would be a reduction of the LOS component [6]. Also, we have computed the Kullback Leibler (KL) distance [15] between the histogram of channel estimate magnitudes obtained from experiments and the probability density function of a Ricean distribution with $K$-factor computed from experiments. The CDFs of the KL distances computed for our experiments are shown in Fig. 6. The low values of KL distances for the Ricean distribution verify that modeling $|h_{\text{I},i}[f]|$ and $|h_{\Phi,i}[f]|$ as Ricean is a good fit. To put these results in more perspective we also show results for the KL distance between the histogram of channel estimate magnitudes and a Rayleigh distribution (which is a Ricean distribution with $K$-factor equal to zero). We observe that Ricean is a better fit than Rayleigh because the KL distances are lower for Ricean.

We first characterize the $K$-factor for the self-interference channel before active cancellation. Fig. 7 shows the CDF of $K_{\text{I},i}$ for our experiments for $d$ equal to 10 cm, 20 cm, and 40 cm. Each of these three CDFs is the result of a total of 16 independent estimates of $K_{\text{I},i}$ at a fixed $d$. Results in Fig. 7 show that the value of $K_{\text{I},i}$ for $d$ between 10 cm and 40 cm is between 25 dB and 40 dB. These large values of $K_{\text{I},i}$ were expected due to the proximity of same node antennas. It is expected that as the distance between antennas increases the value of $K_{\text{I},i}$ decreases. Hence, the CDF of $K_{\text{I},i}$ for $d = 40$ cm should be more to the left than the CDF of $K_{\text{I},i}$ for $d = 20$ cm and the CDF of $K_{\text{I},i}$ for $d = 10$ cm should be more to the right. However, results in Fig. 7 do not show a clear difference between the CDFs for $d = 10$ cm, $d = 20$ cm and $d = 40$ cm. Our intuition is that an increase in separation from $d = 10$ cm to $d = 40$ cm results in a decrease in $K_{\text{I},i}$ that is smaller than the error in our estimate, hence it is not captured by the CDFs shown in Fig. 7. To support this intuition we also show in Fig. 7 the CDF of $K_{\text{S},i}$ which is the estimate of the $K$-factor of the distribution of $|h_{\text{S},i}[f]|$. Since $h_{\text{S},i}[f]$ is the channel between two antennas at LOS placed at distance $D = 8.5$ m the CDF of $K_{\text{S},i}$ should be noticeably to the left of the CDF of $K_{\text{I},i}$ and this is verified by results in Fig. 7. Next, we characterize $K$-factor of the self-interference channel after active cancellation and verify the following result.

*Result 4:* The $K$-factor for the self-interference channel reduces due to active cancellation and the amount of reduction increases as the self-interference cancellation increases. Hence, the $K$-factor for the self-interference channel after active cancellation depends on the $K$-factor value before active cancellation and the suppression achieved by active cancellation.

The CDF of $K_{\text{AC},i}$ and the CDF of $K_{\text{ACDC},i}$ for our experiments are shown in Fig. 7. Results in Fig. 7 show that the $K$-factor before active cancellation is larger than the $K$-factor after active cancellation. Hence, the $K$-factor for the self-interference channel reduces due to active cancellation. This reduction in $K$-factor is a function of the suppression achieved by active cancellation. This is verified by experiment results shown in Fig. 8 where we plot the reduction in $K$-factor due to active cancellation mechanism $\Phi$ (this reduction is computed as $K_{\text{I},i}$ (dB) $-$ $K_{\Phi,i}$ (dB)) as a function of the average amount of suppression $\alpha_{\Phi,i}$ for $\Phi \in \{\text{AC}, \text{ACDC}\}$. As the suppression increases, the reduction in $K$-factor also increases.

*Reasons for Result 4*: Active cancellation is based on estimation of the self-interference channel. Before active cancellation the self-interference channel has a strong LOS component, hence, an estimate of the self-interference channel before active cancellation is virtually an estimate of the strong LOS component of this channel. Consequently, most of the cancellation applied by active cancellation corresponds to attenuation of the strong LOS component that is present before active cancellation and as the suppression achieved by active cancellation increases the attenuation of the LOS component also increases. ■

## V. Achievable Rates

### A. Computation of Achievable Rates

We compute the achievable rate based on the SINR per frame which is estimated based on the Average Error Vector Magnitude Squared (AEVMS) [16] per frame. In our two node experiments the AEVMS per frame transmitted from Node $j$ to Node $i$ is estimated as follows. Symbol $s_j[n,f]$ is sent from Node $j$ to Node $i$ and Node $i$ computes an estimate of $s_j[n,f]$ which we label as $\widehat{s}_j[n,f]$. The AEVMS per frame transmitted from Node $j$ to Node $i$ is estimated as $\text{AEVMS}_i[f] = \frac{1}{N_{\text{sym}}} \sum_{n=1}^{N_{\text{sym}}} |s_j[n,f] - \widehat{s}_j[n,f]|^2$. The SINR for frame $f$ received at Node $i$ is estimated as $\text{SINR}_i[f] = 1/\text{AEVMS}_i[f]$ and the achievable rate for frame $f$ received at Node $i$ is estimated as $\text{AR}_i[f] = \log_2(1 + \text{SINR}_i[f])$. The achievable rate for transmission to Node $i$ is computed by averaging over all the achievable rates per frame and it is equal to $\text{AR}_i = \frac{1}{N_{\text{frames}}} \sum_{f=1}^{N_{\text{frames}}} \log_2(1 + \text{SINR}_i[f])$. The achievable sum rate of the full-duplex two-way system is computed by averaging over all the achievable sum rates per frame and it is equal to $\text{ASR} = \frac{1}{N_{\text{frames}}} \sum_{f=1}^{N_{\text{frames}}} \left(\log_2\left(1 + \text{SINR}_1[f]\right) + \log_2\left(1 + \text{SINR}_2[f]\right)\right)$. In our experiments we used symbols $s_1[n, f]$ and $s_2[n, f]$ that were modulated using QPSK. However, notice that the AEVMS value is independent of the constellation size and shape chosen for symbols $s_1[n, f]$ and $s_2[n, f]$, this was also discussed in [16]. The numerator in the computation of $\text{SINR}_i[f]$ is equal to one because we are using a normalized constellation, if the average energy per symbol was not equal to one then $\text{SINR}_i[f]$ would have to be scaled by a normalization factor.

A natural question that arises is the following. Can the full-duplex systems evaluated in our experiments achieve larger rates than a half-duplex system? In order to answer this question we also ran experiments for a two-way half-duplex $2 \times 1$ Alamouti system [17]. This system uses two antennas, two transmitter radios, and one receiver radio per node, hence, it uses the same antenna and radio resources per node as the full-duplex systems we have implemented. For the two-way half-duplex system the nodes time share the link with 50% of the time dedicated for transmission from Node 1 to Node 2 and 50% of the time dedicated for transmission from Node 2 to Node 1. The achievable sum rate for the half-duplex system is computed as $\text{ASR} = \frac{1}{N_{\text{frames}}} \sum_{f=1}^{N_{\text{frames}}} \left(\frac{1}{2}\log_2\left(1 + \text{SINR}_1[f]\right) + \frac{1}{2}\log_2\left(1 + \text{SINR}_2[f]\right)\right)$.

### *B. Achievable Rates with Increasing Power*

*Result 5:* If the signal to interference ratio before active cancellation at Node $i$, $\text{SIR}_{\text{AS},i}$, is maintained constant while the average received self-interference power at Node $i$, $\text{P}_{\text{RI},i}$, is increased, then the achievable rate for transmission to Node $i$ increases.

Experiment results that verify Result 5 are shown in Fig. 9 where each curve corresponds to an approximately constant value of $\text{SIR}_{\text{AS},i}$. Experiments were performed with both nodes using the same transmission power $\text{P}_{\text{T}}$. The equation for $\text{SIR}_{\text{AS},i}$ can be written as $\text{SIR}_{\text{AS},i}$ (dB) $=$ $\text{P}_{\text{RS},i}$ (dBm) $-$ $\text{P}_{\text{RI},i}$ (dBm), where $\text{P}_{\text{RS},i}$ is the average power of the received signal of interest at Node $i$. Received powers $\text{P}_{\text{RS},i}$ (dBm) and $\text{P}_{\text{RI},i}$ (dBm) are both proportional to $\text{P}_{\text{T}}$. We were able to increase $\text{P}_{\text{RI},i}$, while keeping $\text{SIR}_{\text{AS},i}$ constant, by increasing $\text{P}_{\text{T}}$ at both nodes by the same amount. For each curve in Fig. 9 the majority of the data points show that, although the value of $\text{SIR}_{\text{AS},i}$ is approximately constant, the achievable rate for transmission to Node $i$ is increasing as $\text{P}_{\text{RI},i}$ increases.

*Reasons for Result 5*: Result 5 can be explained using our derived equations and Result 1

as follows. In Result 1 of Section IV-A we showed that $\alpha_{\Phi,i}$ increases as $\mathrm{P}_{\mathrm{RI},i}$ increases. In Section II-E we obtained that the SINR at Node $i$ when using active self-interference cancellation mechanism $\Phi$ is given by $\mathrm{SINR}_{\Phi,i} = 1/(\frac{1}{\alpha_{\Phi,i}\mathrm{SIR}_{\mathrm{AS}_i}} + \frac{1}{\mathrm{SNR}_i})$. Since $\mathrm{SIR}_{\mathrm{AS}_i}$ (dB) $= \mathrm{P}_{\mathrm{RS},i}$ (dBm) $-$ $\mathrm{P}_{\mathrm{RI},i}$ (dBm) we observe the following. If $\mathrm{SIR}_{\mathrm{AS},i}$ remains constant while $\mathrm{P}_{\mathrm{RI},i}$ increases then this means that $\mathrm{P}_{\mathrm{RS},i}$ is increasing and the rate of increase of $\mathrm{P}_{\mathrm{RS},i}$ is the same rate of increase as $\mathrm{P}_{\mathrm{RI},i}$. Hence, if $\mathrm{SIR}_{\mathrm{AS},i}$ remains constant while $\mathrm{P}_{\mathrm{RI},i}$ increases then the terms in the equation for $\mathrm{SINR}_{\Phi,i}$ that are changing are $\alpha_{\Phi,i}$ and $\mathrm{SNR}_i$ and they are both increasing consequently $\mathrm{SINR}_i$ increases and this results in an increase in achievable rate. ■

Notice that if the transmission power at both nodes in a two-way full-duplex system is increased by the same amount then $\mathrm{SIR}_{\mathrm{AS},1}$ and $\mathrm{SIR}_{\mathrm{AS},2}$ will not change and $\mathrm{P}_{\mathrm{RI},1}$ and $\mathrm{P}_{\mathrm{RI},2}$ will increase hence, as can be concluded from Result 5, the achievable rate in both directions of the link will increase. Hence, Result 5 leads to the following design rule for two-way full-duplex.

*Design Rule 1 (Rate-Power Increase)*: In a two way full-duplex system, increasing the transmission power at both nodes by the same amount results in an increase of the achievable rate in both directions of the link, and this increases the achievable sum rate of the full-duplex system.

Experiment results that verify our Design Rule 1 are shown in Fig. 10. All results in Fig. 10 correspond to a distance between nodes equal to 8.5 m and we show results for the three different values of $d$ considered in our experiments. For the full-duplex systems we show results for the case where both nodes use active analog cancellation and active digital cancellation (results labeled as FD-ACDC) and for the case where both nodes use active analog cancellation and do not use digital cancellation (results labeled as FD-AC). The majority of the data points for FD-ACDC and FD-AC in Fig. 10 show that the achievable sum rate of the full-duplex system increases as the transmission power $\mathrm{P}_{\mathrm{T}}$ increases.

Fig. 10 also shows results for half-duplex experiments (results labeled as HD) as a function of the transmit power per antenna which was set equal to $\mathrm{P}_{\mathrm{T}}$. The half-duplex results for different values of $d$ look very similar because all the values of $d$ are greater than half a wavelength. For a fixed distance between nodes the antenna separation at a node will yield the same channel statistics between nodes (independent channels for each transmit-receive antenna pair) for values of $d$ larger than half a wavelength [18]. We observe that for $d = 10$ cm the half-duplex system achieves larger rates than the FD-AC and FD-ACDC systems. For $d = 20$ cm and $d = 40$ cm the full-duplex systems achieve larger rates than the half-duplex system. We conclude that, although a total self-interference cancellation of $\sim$74 dB does not bring the self-interference down to the noise floor, a cancellation of $\sim$74 dB is enough to achieve full-duplex gains over half-duplex at a distance between nodes of 8.5 m. For $d = 10$ cm the total cancellation was equal $\sim$66 dB and we observe that this total cancellation is not enough to achieve full-duplex gains over half-duplex at 8.5 m between nodes.

### C. Achievable Rates with Selective Digital Cancellation

From results in Fig. 5 we observed that there are frames where applying digital cancellation after analog cancellation reduces the self-interference ($\alpha_{\text{DC},i}[f] > 0$) but there are also frames where it increases the self-interference ($\alpha_{\text{DC},i}[f] < 0$). We would like to apply digital cancellation only during frames where it reduces the self-interference. Notice that training signals can be used to estimate $\alpha_{\text{DC},i}[f]$. Using $\alpha_{\text{DC},i}[f]$ estimated based on training, Node $i$ can decide if digital cancellation should be applied to frame $f$ based on the following design rule.

*Design Rule 2 (Selective Digital Cancellation)*: If $\alpha_{\text{DC},i}[f] > 0$ then apply digital cancellation to the payload received at Node $i$ during frame $f$, otherwise do not apply digital cancellation to the payload received at Node $i$ during frame $f$.

We performed experiments for a full-duplex system with analog cancellation and the frames received were stored and post-processed in the following three different ways. (1) We did not apply digital cancellation to the received payload and this payload was used to compute the achievable rate of a Full-Duplex system with Analog Cancellation (FD-AC). (2) We applied digital cancellation to the received payload of all frames and the resulting payload was used to compute the achievable rate of a Full-Duplex system with Analog Cancellation and Digital Cancellation (FD-ACDC). (3) At each node we applied digital cancellation to the received payload selectively based on Design Rule 2 and the resulting payload was used to compute the achievable rate of a Full-Duplex system with Analog Cancellation and Selective Digital Cancellation (FD-ACSDC). Fig. 11 shows achievable rate results for these three full-duplex systems. Each thick bar in Fig. 11 shows the achievable rate averaged over all frames. The lower end of each thin bar in Fig. 11 indicates the lowest achievable rate per frame obtained for the corresponding full-duplex system and the upper end of each thin bar indicates the largest achievable rate per frame obtained for the corresponding full duplex system. Hence, thin bars show the range of achievable rate values observed for each full-duplex system and thick bars show the average over all the achievable rate values per frame.

By comparing the thick bars in Fig. 11, we conclude that the best full-duplex system is the one that applies digital cancellation selectively based on Design Rule 2. By comparing the lower end of the thin bars in Fig. 11, we observe that the lowest achievable rate per frame tends to be higher for the systems that use digital cancellation. The reason is that for the frames where analog cancellation has poor performance (leading to a low achievable rate of the FD-AC system for those frames) digital cancellation can increase the total self-interference suppression and this leads to a larger achievable rate for the FD-ACDC and FD-ACSDC systems for those frames. Hence, digital cancellation is an excellent "safety net" for the frames where analog cancellation delivers poor suppression. A comparison of the upper end of the thin bars in Fig. 11 shows that applying digital cancellation all the time can sometimes result in a reduction of the achievable rate but applying digital cancellation selectively based on Design Rule 2 avoids applying digital cancellation in frames where it would result in a decrease in achievable rate.

## VI. Conclusions

In order to advance the theory of full-duplex systems it is important to have signal models based on actual measurements. We contribute to this area by providing a characterization of the distribution of the self-interference before and after active cancellation mechanisms. We are the first ones to report a statistical characterization of the self-interference based on extensive measurements.

We have characterized the effect of increasing self-interference and transmission power on the rate and cancellation performance of our full-duplex implementation. Such a characterization, while fundamental for the future deployment of full-duplex systems, is seldom provided in related work on full-duplex implementation.

We are the first to characterize the performance of digital cancellation as a function of the performance of analog cancellation for a full-duplex implementation. The current belief in the literature has been that digital cancellation will always help improve the total cancellation. Further, it has some times been conjectured that the total analog plus digital cancellation would be equal to the sum of the cancellations measured independently in isolation. We demonstrate that when digital cancellation is preceded by analog cancellation, the amount of digital cancellation varies as a function of the amount of analog cancellation. Thus, only a full-system implementation can reveal the true benefits of combined analog and digital cancellation in an actual system.

Finally, while our results are based on a single implementation of one full-duplex system, we believe our results analyze fundamental characteristics of full-duplex systems that had not been analyzed before.

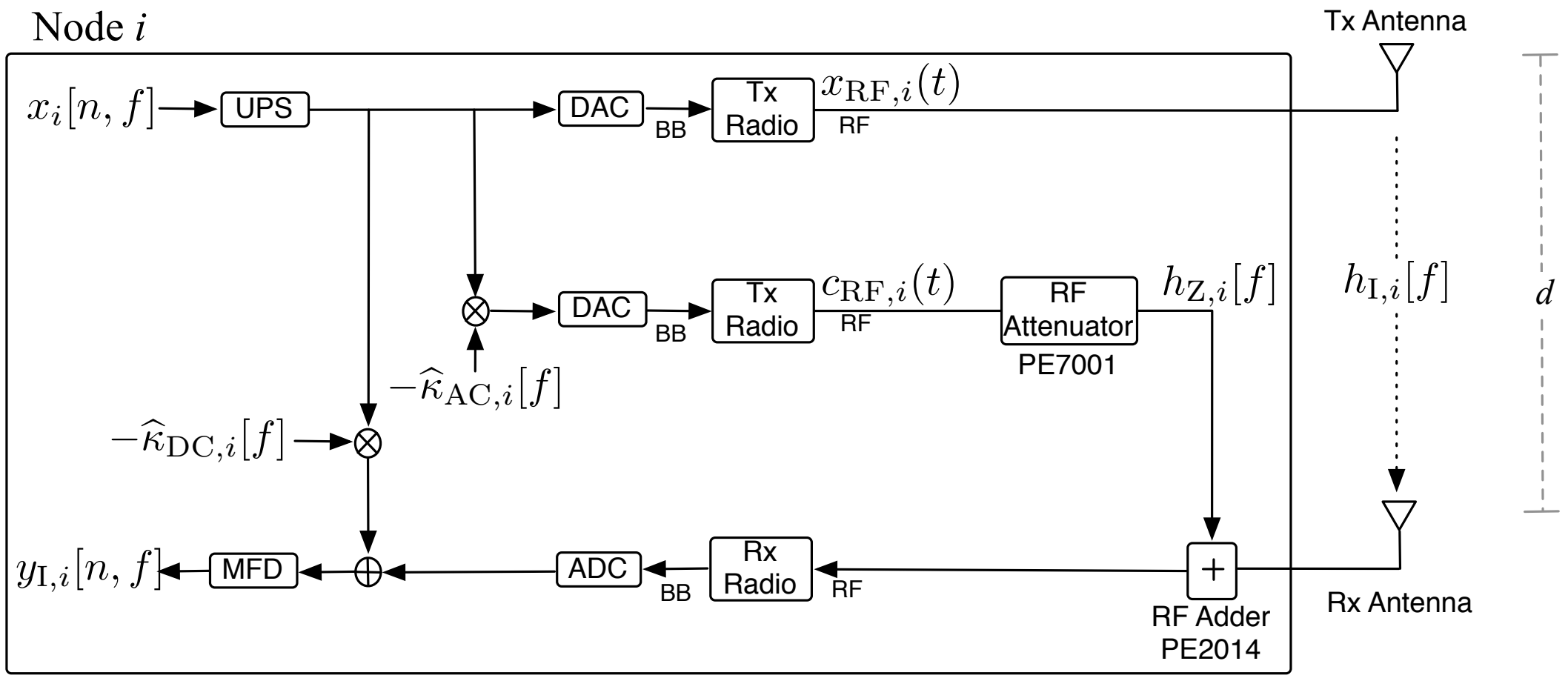

Fig. 1. Block diagram of a full-duplex node.

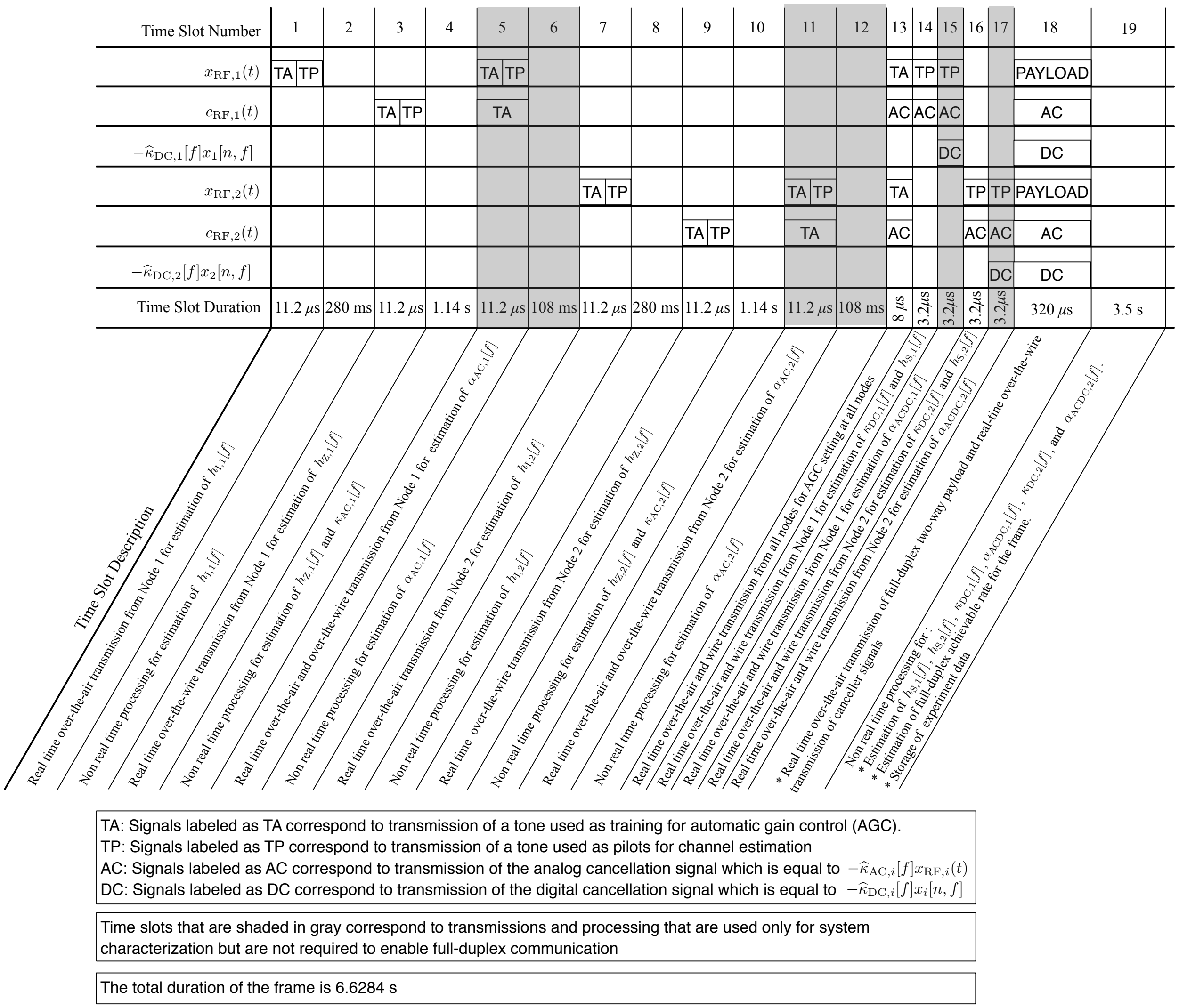

Fig. 2. Time diagram for a full-duplex frame.

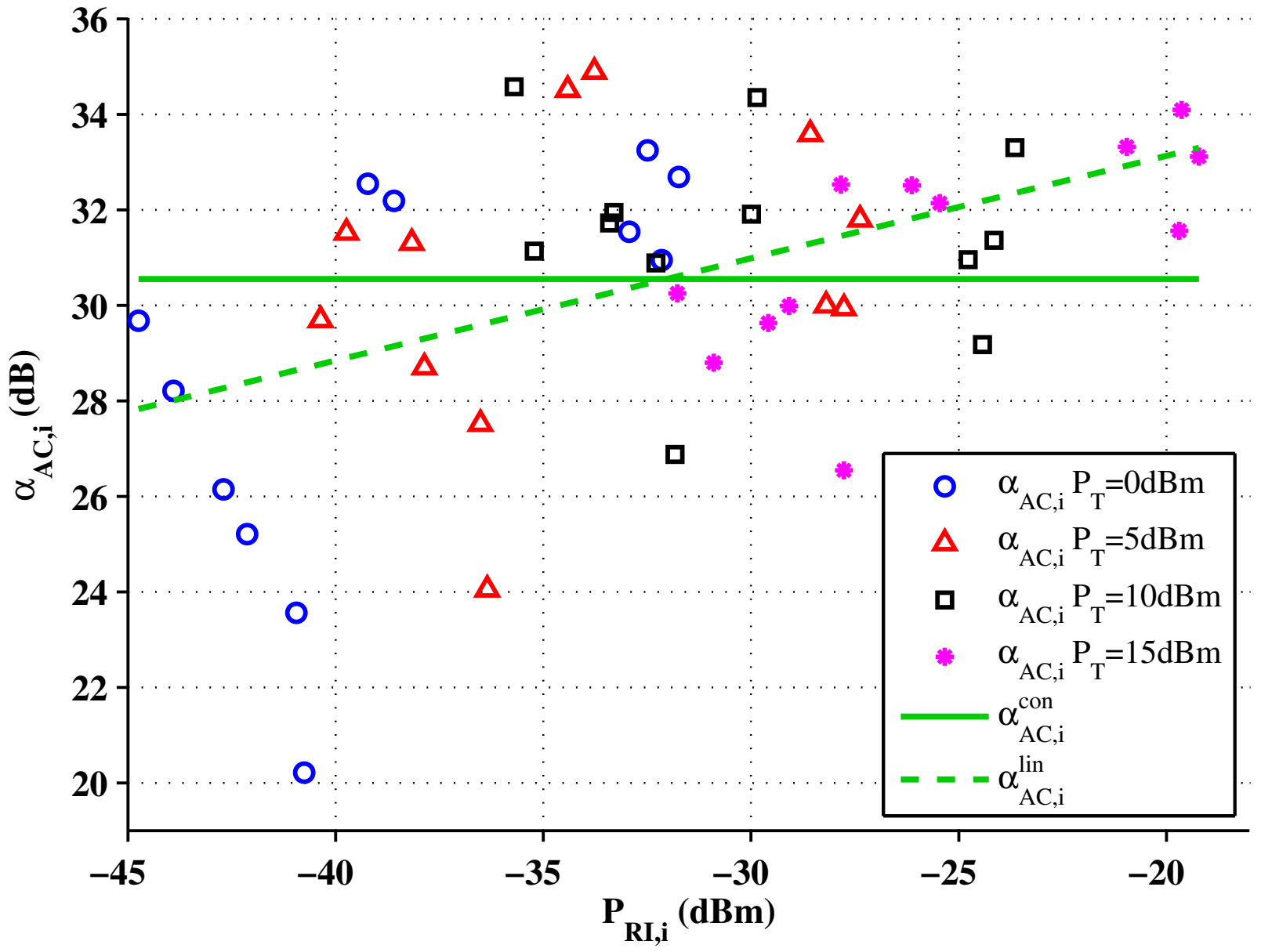

(a) Results for active analog cancellation. The constant fit is equal to $\alpha^{\text{con}}_{\text{AC},i} = 30.55$ dB and the linear fit is equal to $\alpha^{\text{lin}}_{\text{AC},i} = \lambda_{\text{AC}}\text{P}_{\text{RI},i} + \beta_{\text{AC}}$ where $\lambda_{\text{AC}} = 0.21$ dB/dBm and $\beta_{\text{AC}} = 37.42$ dB.

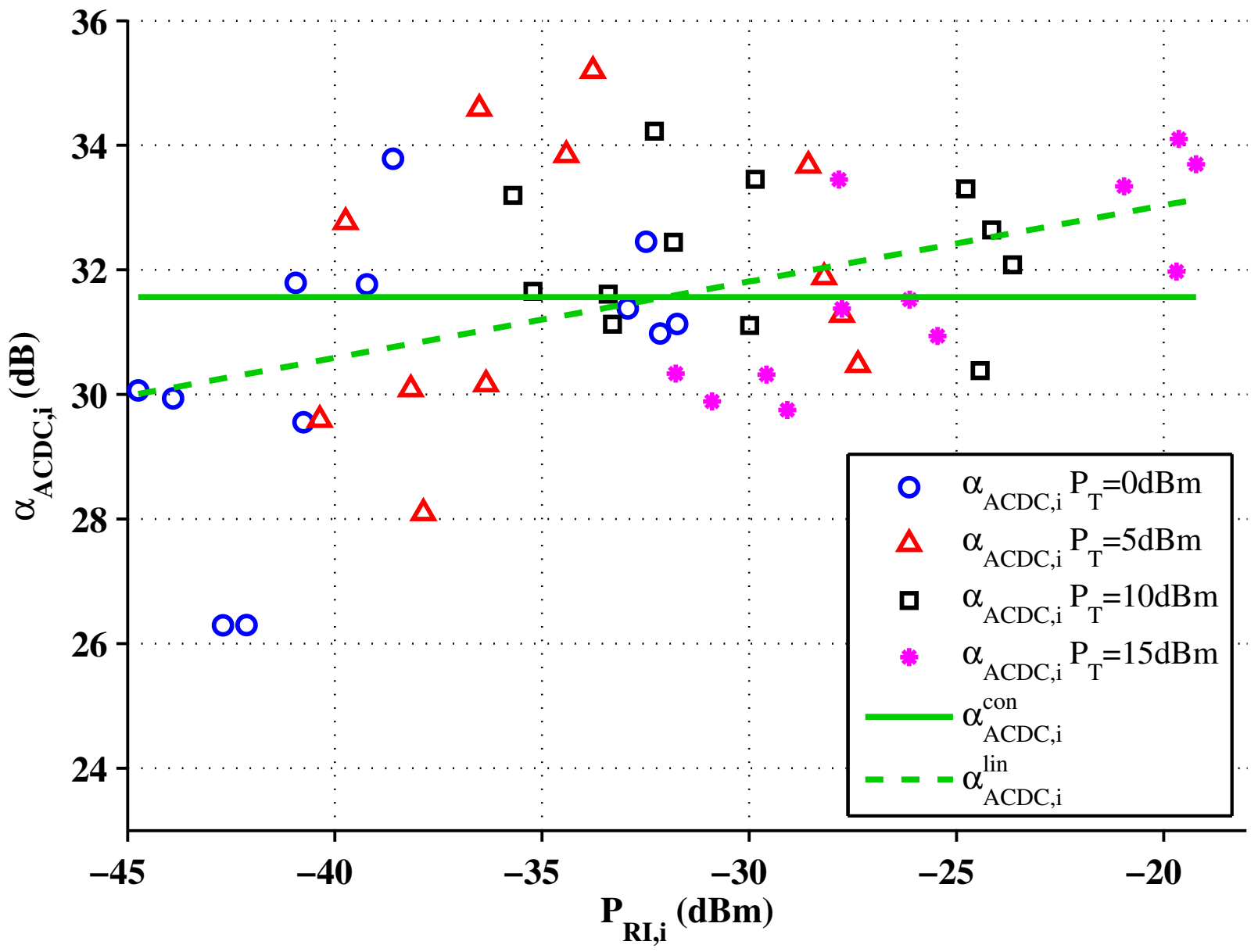

(b) Results for combined active analog and active digital cancellation. The constant fit is equal to $\alpha^{\text{con}}_{\text{ACDC},i} = 31.56$ dB and the linear fit is equal to $\alpha^{\text{lin}}_{\text{ACDC},i} = \lambda_{\text{ACDC}}\text{P}_{\text{RI},i} + \beta_{\text{ACDC}}$ where $\lambda_{\text{ACDC}} = 0.12$ dB/dBm and $\beta_{\text{AC}} = 35.49$ dB.

Fig. 3. Measurements of the average amount of active cancellation achieved and constant and linear fit for the measurements.

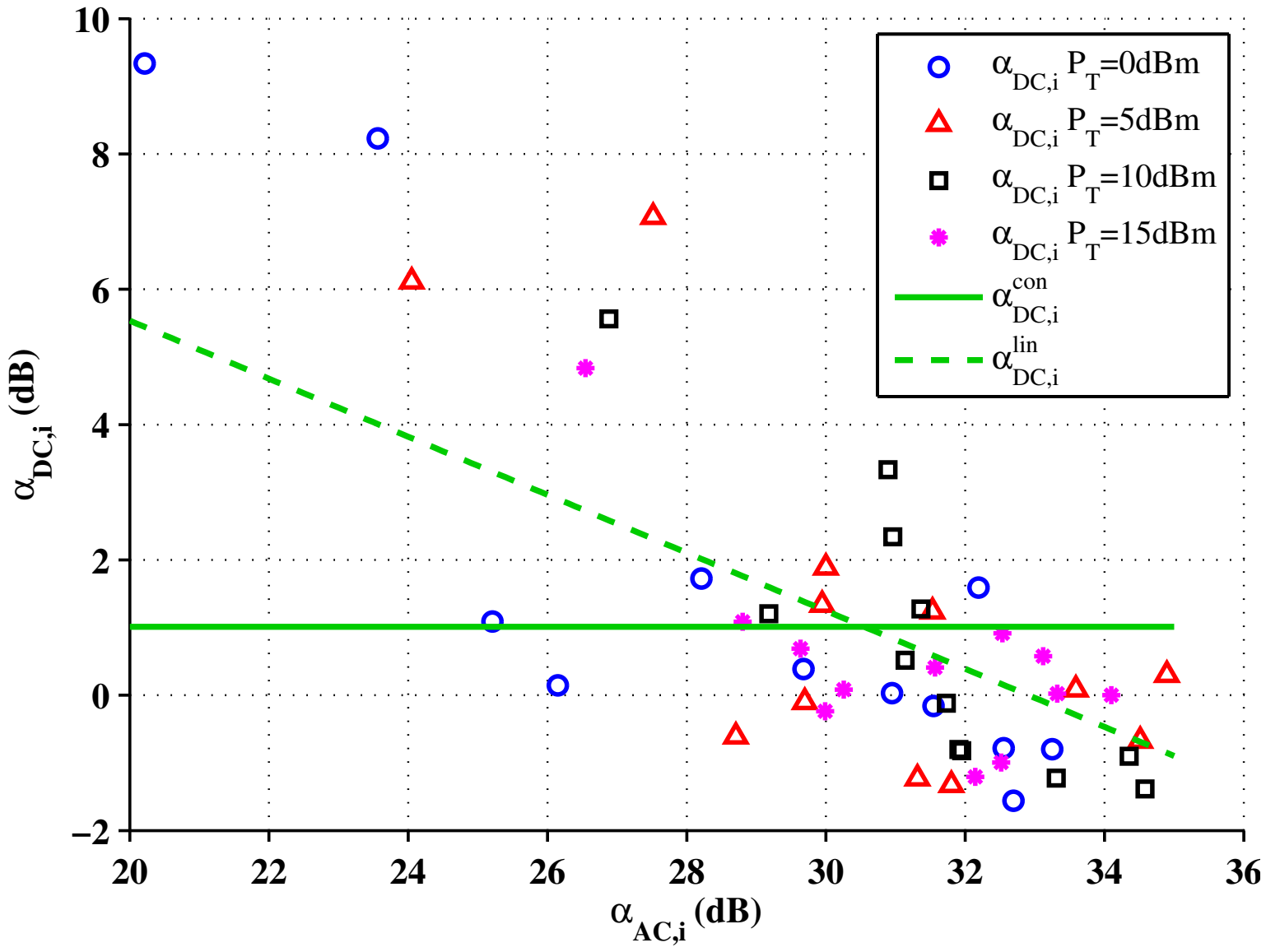

Fig. 4. Measurements of the average amount of cancellation achieved by digital cancellation and cancellation values computed based on the constant and linear fit. The constant fit is equal to $\alpha^{\text{con}}_{\text{DC},i}$ (dB) $= \alpha^{\text{con}}_{\text{ACDC},i} - \alpha^{\text{con}}_{\text{AC},i} = 1.1$ (dB). We compute the linear fit using the equations for $\alpha^{\text{lin}}_{\text{AC},i}$ and $\alpha^{\text{lin}}_{\text{ACDC},i}$. The linear fit is equal to $\alpha^{\text{lin}}_{\text{DC},i}$ (dB) $= \alpha^{\text{lin}}_{\text{ACDC},i}$ (dB) $- \alpha^{\text{lin}}_{\text{AC},i}$ (dB) $= \lambda_{\text{ACDC}}(\alpha^{\text{lin}}_{\text{AC},i}\text{ (dB)} - \beta_{\text{AC}})/\lambda_{\text{AC}} + \beta_{\text{ACDC}} - \alpha^{\text{lin}}_{\text{AC},i}$ (dB)

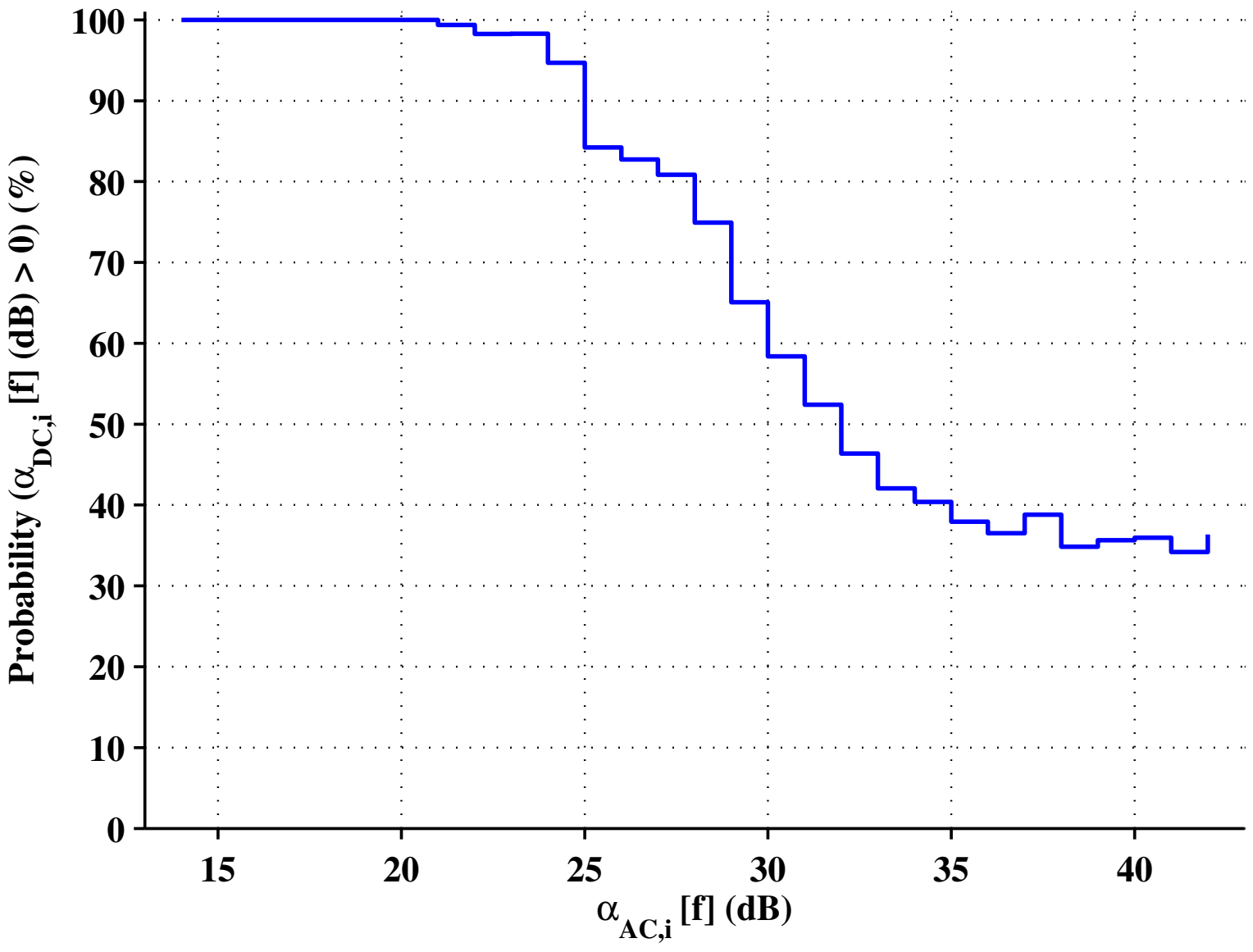

Fig. 5. Probability that digital cancellation after analog cancellation increases the total amount of cancellation during a frame as a function the cancellation achieved by analog cancellation

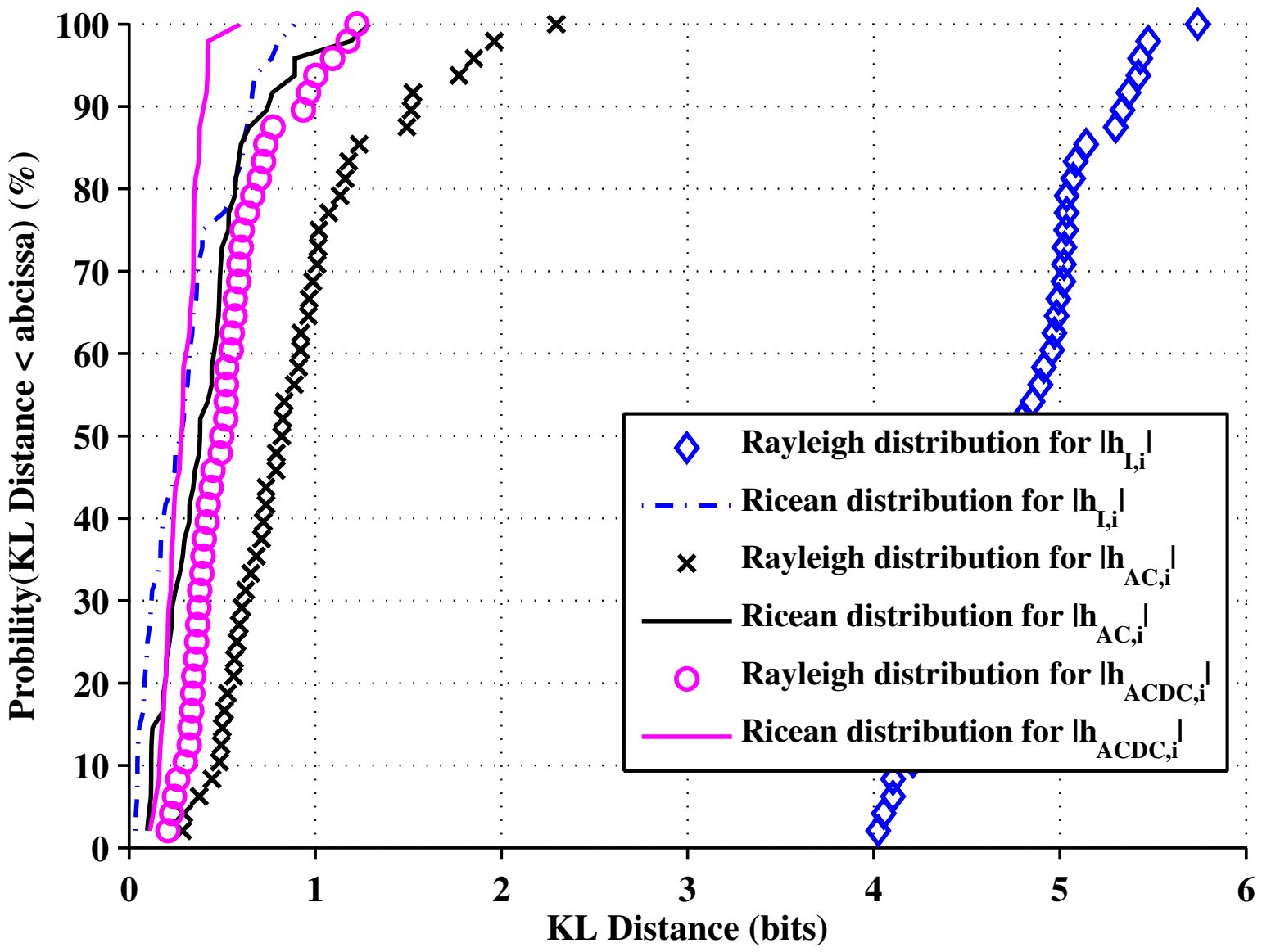

Fig. 6. CDF of KL distance.

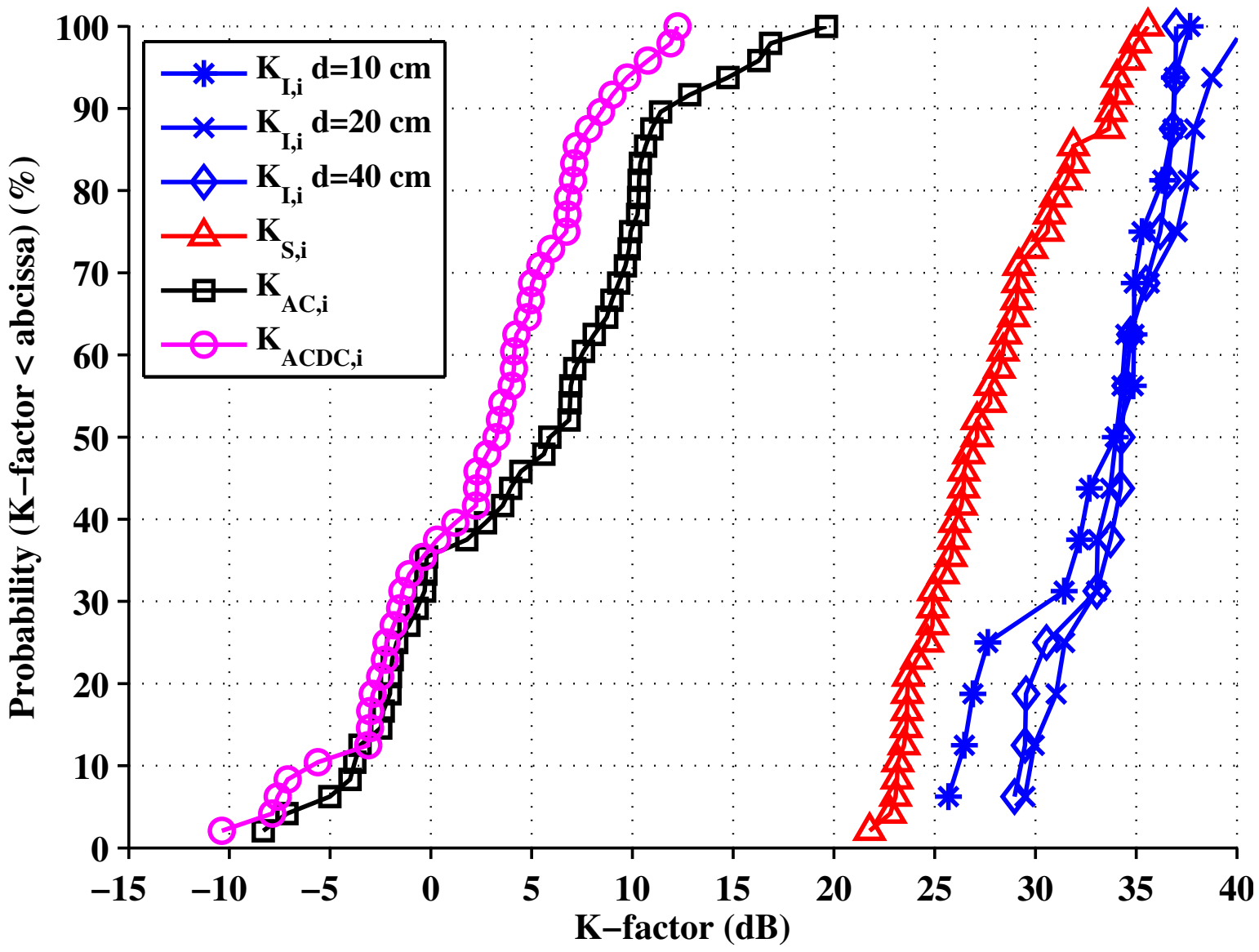

Fig. 7. CDF of $K$-factor. We aggregate $K_{\mathrm{I},i}$ values computed for $i = 1, 2$ and different values of $\mathrm{P_T}$ to obtain the CDF of $K_{\mathrm{I},i}$ for a fixed $d$. We obtained the CDF of $K_{\mathrm{AC},i}$ and $K_{\mathrm{ACDC},i}$ by aggregating the $K$-factors computed for $i = 1, 2$ for different values of $d$ and $\mathrm{P_T}$.

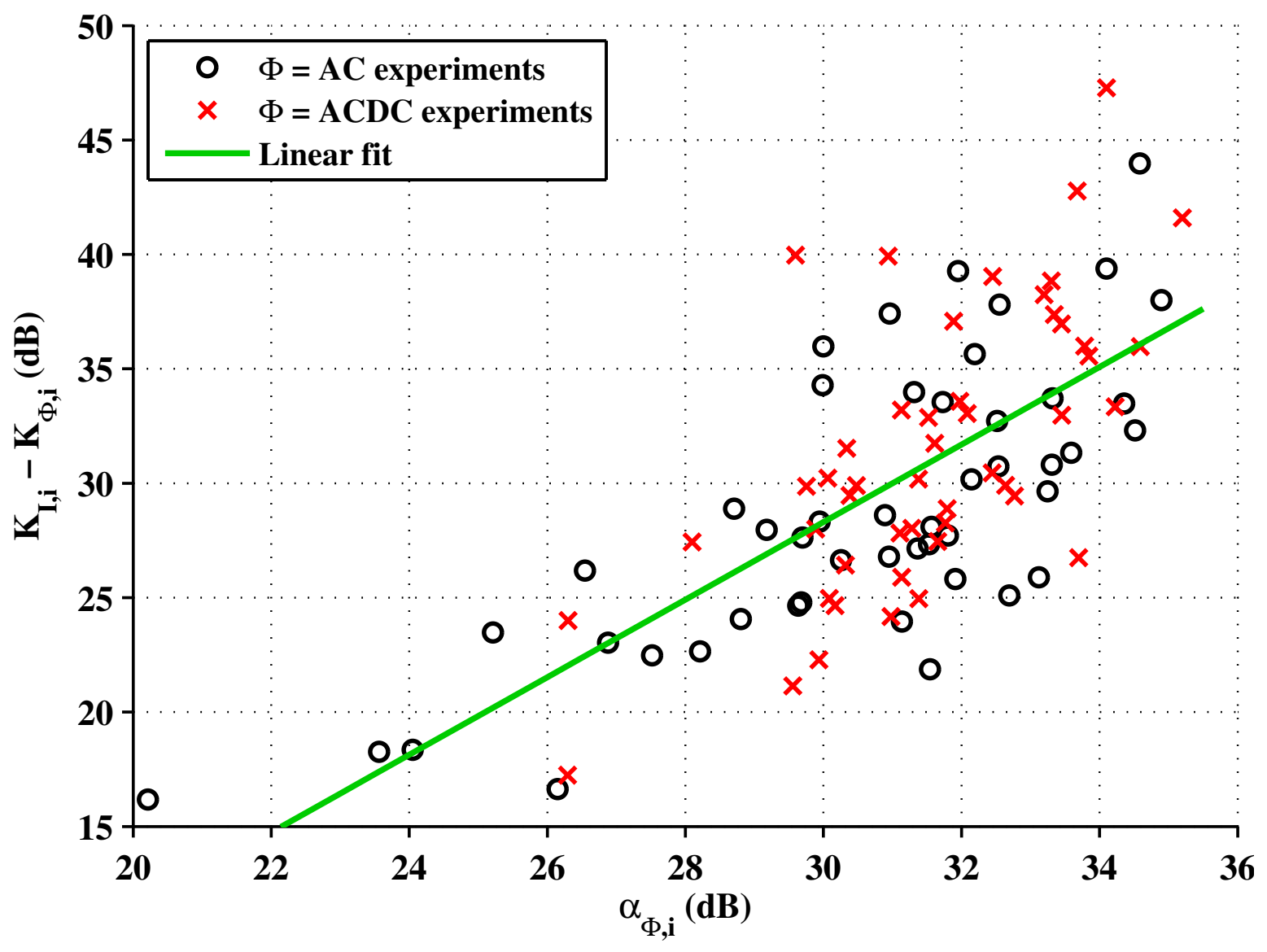

Fig. 8. Reduction in $K$-factor as a function of the average amount of cancellation.

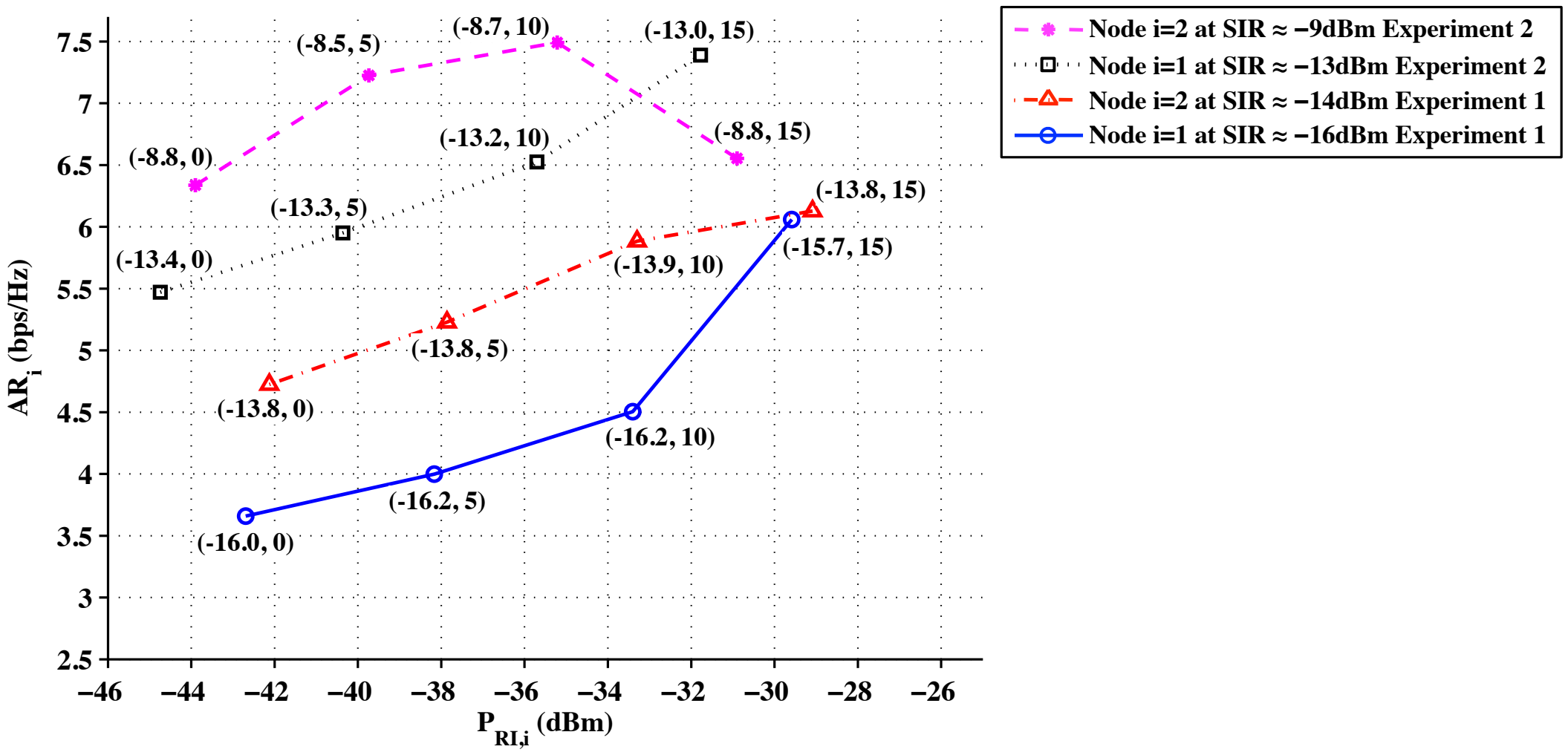

Fig. 9. Experiment results showing the achievable rate for transmission to Node $i$ as a function of the average received self-interference power at Node $i$ for an approximately constant value of $\mathrm{SIR}_{\mathrm{AS},i}$. All the results correspond to a full-duplex system with active analog cancellation, active digital cancellation, a distance $d = 40$ cm between self-interfering antennas, and a distance of 8.5 m between Node 1 and Node 2. Both nodes used the same transmission power $\mathrm{P_T}$ and we were able to increase $\mathrm{P}_{\mathrm{RI},i}$, while keeping $\mathrm{SIR}_{\mathrm{AS},i}$ constant, by increasing $\mathrm{P_T}$. The exact values of $\mathrm{SIR}_{\mathrm{AS},i}$ and $\mathrm{P_T}$ for each data point are shown in parenthesis as a pair of values $(\mathrm{SIR}_{\mathrm{AS},i}\ (\mathrm{dB}), \mathrm{P_T}\ (\mathrm{dBm}))$ next to each data point. Each data point corresponds to an 800 frame experiment and for each node we show two curves because we did the same 800 frame experiment twice. The second experiment was done a few days after the first experiment hence the conditions surrounding the setup were not exactly the same for the two experiments (due movement of people in the laboratory) and this explains why the rates at a node are different between experiments.

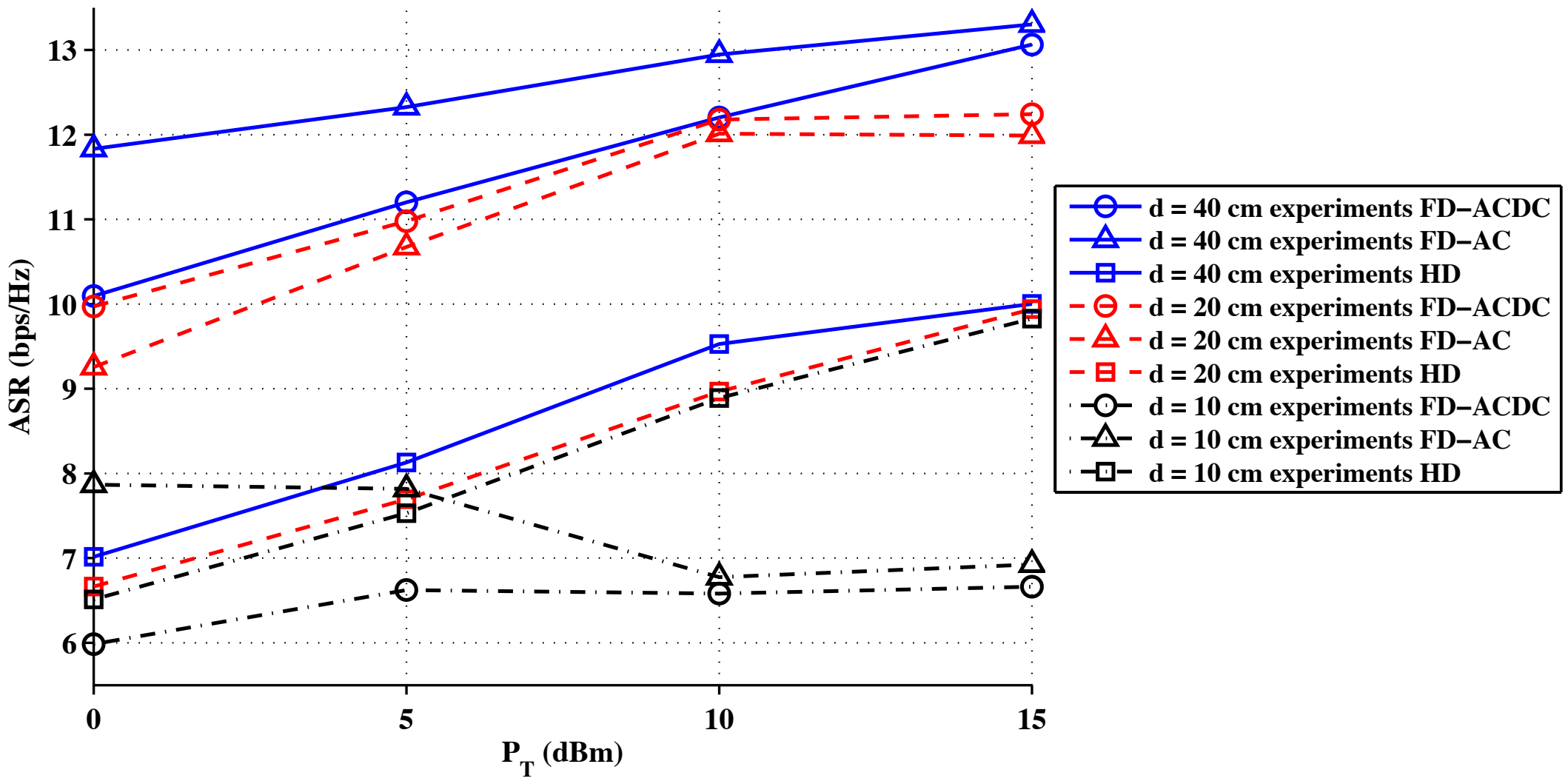

Fig. 10. Experiment results showing an increase in the achievable sum rate as a function of the transmission power for a full-duplex two-way system using the same transmission power $P_T$ at both nodes. The figure also shows results for half-duplex experiments where the transmit power per antenna was set equal to $P_T$. Hence, for each $P_T$ value, the systems compared have the same average transmitted power per node. Each data point corresponds to two 800 frame experiments whose results were averaged to obtain one data point.

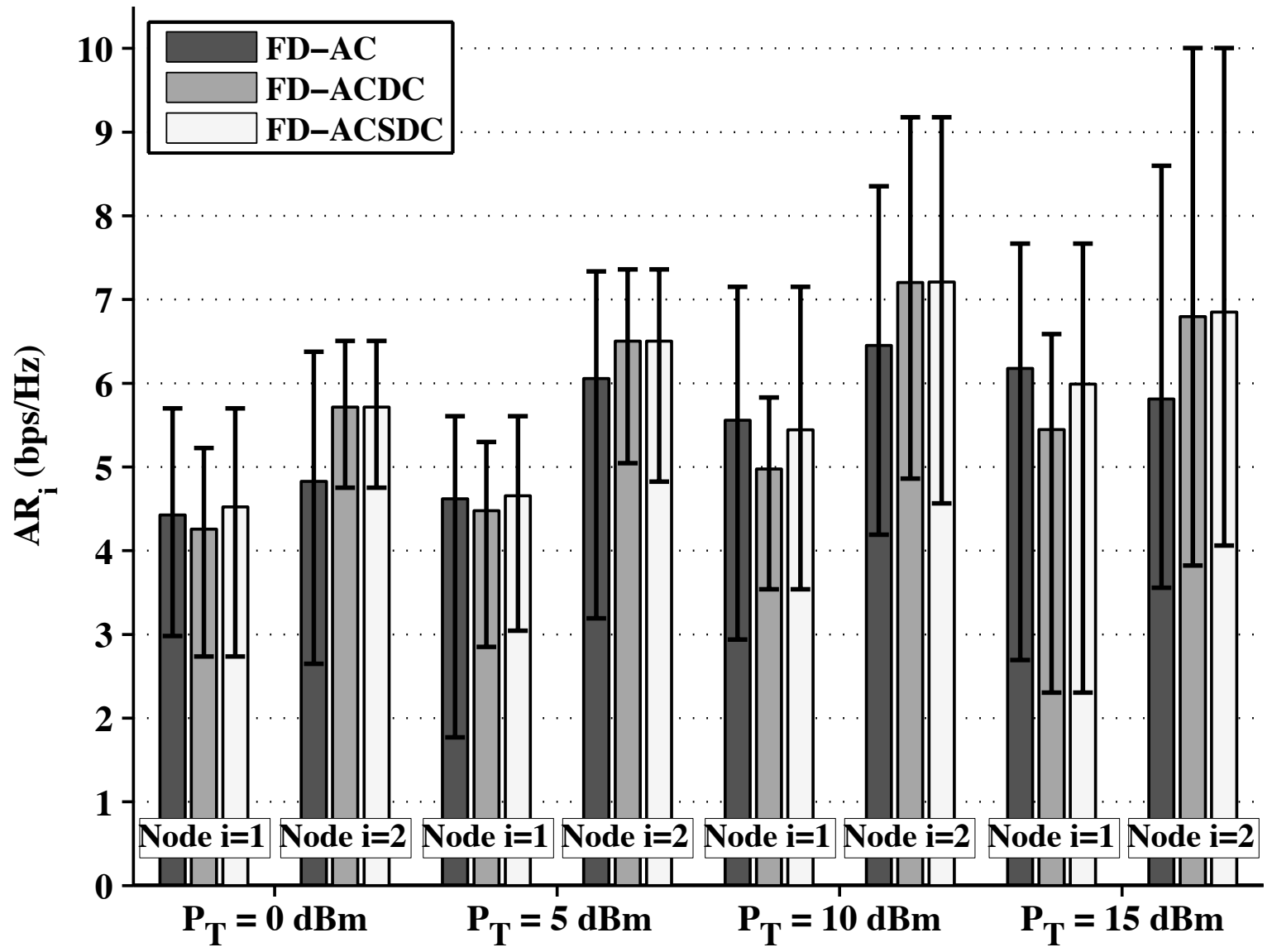

Fig. 11. Experiment results showing the effect of different cancellation schemes on the achievable rate per node of a full-duplex two-way system using the same transmission power $P_T$ at both nodes. Results correspond to $d = 20$ cm and a distance between Node 1 and Node 2 equal to 8.5 m. Each bar corresponds to results from two 800 frame experiments.